\documentclass[twocolumn,epsfig,floats,aps]{revtex4}
\usepackage{amsmath,bm}
\usepackage{epsfig}
\usepackage{graphicx}

\begin{document}
\title{Correlation between event multiplicity
              and elliptic flow parameter}
\author{Dai-Mei Zhou$^1$\footnote{zhoudm@phy.ccnu.edu.cn},
        Yun Cheng$^{1,2,5}$,
        Yu-Liang Yan$^3$,
        Bao-Guo Dong$^3$,
        Ben-Hao Sa$^{1,3,4}$\footnote{sabh@ciae.ac.cn},
        Laszlo P. Csernai$^{2,5}$\footnote{csernai@ift.uib.no}}

\affiliation{ $^1$  Institute of Particle Physics, Huazhong Normal
      University,Wuhan, 430082 China \\
$^2$  Institute of Physics and Technology, University of Bergen,
       Allegaten 55, 5007 Bergen, Norway \\
$^3$  China Institute of Atomic Energy, P. O. Box 275 (18),
      Beijing, 102413 China \\
$^4$  CCAST (World Lab.), P. O. Box 8730 Beijing, 100080 China\\
$^5$  Frankfurt Institute for Advanced Studies,
      Johann Wolfgang Goethe University,
       Ruth-Moufang-Str. 1,
       60438 Frankfurt am Main,
       Germany}

\begin{abstract}
The elliptic flow parameter (component), $v_2$, in the Fourier
expansion of event-by-event charged particle multiplicity
azimuthal distribution in the momentum space is studied by taking
into account the multiplicity fluctuations. The correlations
between charge multiplicity and impact parameter as well as
between elliptic flow parameter and impact parameter
(multiplicity) are investigated by both the parton and hadron
cascade model PACIAE and a multiple phase transport model AMPT. A
discussion is given for the event-wise average and particle-wise
average in the definition of elliptic flow parameter.
\end{abstract}
\maketitle
\section{INTRODUCTION}
The charge particles emitted from the fireball created in
relativistic nucleus-nucleus collisions exhibit transverse
collective flow. This is represented by the elliptic flow parameter
($v_2$) and other harmonic parameters ($v_n$, $n$=0, 1, 3, 4, ...).
Those parameters, as Fourier expansion coefficients of produced
particle azimuthal distribution in the momentum space, are highly
sensitive to the spatial geometry (eccentricity) of the created
fireball (nuclear overlap region or interaction region).

The expected phase transition to Quark-Gluon-Plasma (QGP) should
have a dramatic effect on those harmonic parameters. The
consistency between experimental data of $v_2(p_T)$ ( $v_2(y)$) at
mid-rapidity and the corresponding hydrodynamic predictions is
regarded as an evidence of the production of partonic matter in
the ultra-relativistic nucleus-nucleus collisions
\cite{miclo,csr}. The elliptic flow of high $p_T$ particles may be
related to jet fragmentation and parton energy loss \cite{wang1},
which are usually not included in the hydrodynamic calculations.
Such kind of hydrodynamic calculation \cite{kolb} overestimates
$v_2(p_T)$ in the $p_T \geq 1.5$ GeV/c region \cite{phob1}. This
is regarded as an evidence of the strongly coupled QGP formation
in the relativistic nucleus-nucleus collisions together with the
discovery of jet quenching \cite{zajc}. So far a lot of
experimental data have been published on the collective flow
parameters \cite{star1,phen1,phob2}. Consequently microscopic
transport model studies are also widely progressing
\cite{bin1,fuchs,chen1,xu1,zhu} as well as the abundant
hydrodynamic investigations.

According to two well known pioneering works in this field
\cite{zhang,posk}, the usual study starts from the triple differential
distribution
\begin{equation}
E{\frac{d^3N}{d^3p}}=\frac{1}{2\pi}{\frac{d^2N}{p_Tdydp_T}}
       [1+\sum_{n=1,...}2v_ncos[n(\phi-\Psi_r)]] \label{eq14},
\end{equation}
where $N$ is the particle multiplicity distribution, $\phi$ stands
for the azimuthal angle of particle, and $\Psi_r$ refers to the
azimuthal angle of reaction plane in the momentum space. Then the
n-th flow harmonics is defined as $v_n=\langle
cos[n(\phi-\Psi_r)]\rangle$, where $\langle \rangle$ indicates an
average over all particles in all events of a sample. For the
distribution of all particles in the sample, the coefficients in
Eq. (\ref{eq14}) can be calculated as $v_1 = \langle p_x/p_T
\rangle$ and $v_2 = \langle (p_x/p_T)^2-(p_y/p_T)^2 \rangle$, etc.
This kind of average is widely accepted and will be indicated as
particle-wise average hereafter.

The problem of the particle-wise average in a sample with wider
multiplicity range is that it does not take the influence of
multiplicity (hence impact parameter) in a single event of the
sample into account. In \cite{sazh1} the elliptic flow and other
harmonics have been re-derived starting from the invariant
particle multiplicity distribution in the momentum space. It
turned out that the harmonic parameters $v_n$ (elliptic flow
parameter $v_2$) is an event-wise average of $cos(n\phi)$
\begin{equation}
v_n^e=\langle \overline{cos(n\phi)}\rangle_{ev}{\hspace{0.2cm}
(n=1, 2, ...),} \label{eq3}
\end{equation}
where $\overline{cos(n\phi)}$ denotes the average of $cos(n\phi)$
over particles in a single event, $\langle...\rangle_{ev}$ means
an average over events in a sample, and the superscript ``$e$''
stands for the event-wise average. It is an average of
$cos(n\phi)$ first over particles in a single event then over the
events in a sample. Here it has to mention that in theory if the
beam direction and impact parameter vectors are fixed at the $p_z$
and $p_x$ axes, respectively, then the reaction plane is just the
$p_x-p_z$ plane \cite{zhang}. Therefore the reaction plane
azimuthal angle ($\Psi_r$) in Eq. (\ref{eq14}) introduced for the
extraction of elliptic flow in experiments \cite{posk} is zero.
Meanwhile, the particle-wise average of $v_n$ is also derived in
\cite{sazh1}
\begin{equation}
v_n^p=\left\langle \overline{cos(n\phi)} N_{ev}\right\rangle_{ev}\ /\
\left\langle N_{ev} \right\rangle_{ev}.
\end{equation}
That is obviously different from event-wise average. Only if the
$\overline{cos(n\phi)}$ is independent of event multiplicity
$N_{ev}$ (i. e. if the multiplicity plays no role in the average)
the $v_n^p$ reduces to $v_n^e$. In fact, the
$\overline{cos(n\phi)}$ and $N_{ev}$ correlate (even negatively
correlate) with each other. This is because larger event
multiplicity arises from more central collisions (larger overlap
region between colliding nuclei) and the larger overlap region, in
turn, results in less azimuthal asymmetry. The particle-wise
average does not take the influence of event multiplicity into
account, thus it is questionable from physics point of view. Of
course, for a very narrow multiplicity bin studied, the
particle-wise average is not very problematic. However for the
wide multiplicity bin the correction is important. Here we do not
take the influence of eccentricity fluctuation on $v_2$ into
account, it may strengthen the importance of the correction.

In experiment, the reaction plane is different event by event. In
order to extract the elliptic flow parameter one has to invoke a
complex reaction plane identification method \cite{posk}, the
cumulant method \cite{borg}, and the Lee-Yang zeroes method
\cite{lee}. In all of these methods a quantity has to be first
constructed event by event. This quantity is just the event plane
in \cite{posk}, the cumulant expansion of the weighted $n$-$th$
transverse event-flow vector in \cite{borg}, and a generating
function in \cite{lee}. Then a corresponding average over measured
events has to be taken. Therefore the experimental extraction of
elliptic flow parameter is an event-wise average. In the recent
paper \cite{casa} it has been pointed out that ``Elliptic flow
develops on an event-by-event basis. However, the experimental
determination of $v_2$ demands averaging over events and thus over
distributions of geometric shapes." This means again the
experimental extraction of $v_n$ is an event-wise average.

In the expansion process of a nucleus-nucleus collision, the
initial spatial asymmetry of the created fireball evolves into an
azimuthal asymmetry ($v_n$) in the transverse momentum
distribution of produced charged particles \cite{phob2}. This
evolution process has to be described precisely by detailed
dynamical models. Thus, in this work a parton and hadron cascade
model PACIAE \cite{sa2} and a multiple phase transport model AMPT
(with string melting) \cite{ampt1} are both used in order to have
cross checking. The observables of the impact parameter ($b$),
charged particle multiplicity ($N_{ch}$), eccentricity
($\epsilon$), and elliptic flow parameter ($v_2$) as well as other
harmonic parameters ($v_n$) are always used to describe the
physics of the initial spatial asymmetry, final momentum
asymmetry, and their fluctuations and correlations. As first step,
in this paper we only study the correlation between $N_{ch}$ and
$b$ as well as the correlation between $v_2$ and $b$ ($N_{ch}$).
In addition, the event-wise average versus particle-wise average
in calculating (extracting) the elliptic flow parameter raised in
\cite{sazh1} is also studied. The eccentricity fluctuation may
have extra effect on the $v_2$ parameter and its fluctuation
besides the impact parameter (multiplicity) fluctuation, which
will be studied in next step. Our studies are based on the
multiplicity fluctuation at fixed impact parameter, and would
apply also if the initial eccentricity would not fluctuate.

\vspace{0.4in}
\begin{figure}[ht]
\epsfig{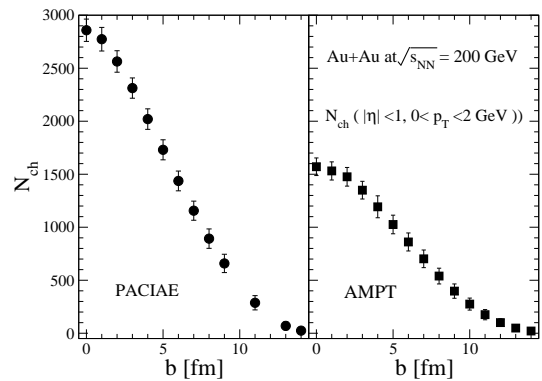}
\vspace{-0.1in} \caption{Correlation between impact parameter $b$
and charge multiplicity $N_{ch}$ calculated by the PACIAE model
(left panel) and the AMPT model (right panel) for the indicated
pseudo-rapidity and $p_T$ range. The error bars indicate the
fluctuations of multiplicity at fixed impact parameter.
Consequently a fixed event multiplicity may correspond to
different impact parameters.} \label{multib}
\end{figure}

\section{MODELS}
The PACIAE model \cite{sa2} is a parton and hadron cascade model and
consists of four stages: parton initialization, parton evolution
(rescattering), hadronization, and hadron evolution (rescattering).

In the first stage, a nucleus-nucleus collision is decomposed into
nucleon-nucleon (NN) collisions according to the collision geometry:
\begin{itemize}
\item Both colliding nuclei settle down at the distance between
two centers of colliding nuclei along the x-axis is equal to
impact parameter $b$. \item Nucleons in the colliding nucleus are
randomly distributed around the center of nucleus according to a
Wood-Saxon distribution ($r$) and the 4$\pi$ uniform distribution
($\theta$ and $\phi$). \item The beam momentum is given to
nucleons in the colliding nucleus. The nucleon Fermi motion is
neglected. \item The NN collision is sampled randomly according to
the spatial and momentum distributions, the straight line
trajectories, and the NN interaction cross sections of the
colliding pairs.
\end{itemize}
Then a NN collision is described by the PYTHIA model \cite{soj2}. In the
PYTHIA model
\begin{itemize}
\item The NN collision is decomposed into parton-parton
collisions. \item A hard parton-parton collision is described by
the lowest leading order perturbative QCD (LO-pQCD) parton-parton
cross sections modified by the parton distribution function in a
nucleon. \item The soft parton-parton collision is considered
empirically. \item Because the initial- and final-state QCD
radiations are considered in parton-parton interactions the
consequence of a NN collision is partonic multijet configuration
composed of di-quarks (anti-diquarks), quarks (anti-quarks),
gluons, and a few hadronic remnants. \item The string
fragmentation is performed. Then one obtains a final hadronic
state for a NN collision (hence for a nucleus-nucleus collision).
\end{itemize}
However, in the PACIAE model the above string fragmentation is
switched-off temporarily. The di-quarks (anti-diquarks) are broken
randomly into quarks (anti-quarks). So, the consequence of a NN
collision (hence a nucleus-nucleus collision) is a configuration
of quarks, anti-quarks, gluons, and hadronic remnants. In
addition, there are spectator nucleons for a nucleus-nucleus
collision.

The partonic rescattering stage follows the initialization one. In
this stage the rescattering among partons are considered by the $2
\rightarrow 2$ LO-pQCD differential cross sections \cite{comb}.
The six elastic and three inelastic parton-parton scatterings are
involved with their own differential cross sections \cite{comb}.
We use the Monte Carlo method to simulate the parton rescattering
until the parton-parton collisions cease.

In the hadronization stage the partons after rescattering are
hadronized by either the string fragmentation scheme (Lund string
fragmentation and/or independent fragmentation) \cite{soj2} or the
coalescence model \cite{sa2}.

The last stage of hadronic rescattering is described by the usual two-body
collision method \cite{sa1}. As such, the hadronic matter is evolved until
the hadron-hadron collision pairs are exhausted.

The two models of PACIAE and AMPT with string melting are physically
similar.
They have similar model assumptions but have otherwise very little commons
in detail. The main differences between them are as follows:
\begin{itemize}
\item AMPT is based on HIJING \cite{wang1} instead of PYTHIA in
PACIAE. \item In the AMPT model the hadrons from HIJING are
fragmented into partons, while the spectator nucleons are kept
survival. This parton initialization for a nucleus-nucleus
collision is quite different from the one in the PACIAE model,
where parton initialization is realized by broking the strings
before the string fragmentation. Thus, even HIJING itself is
heavily based on PYTHIA, the initialized parton state is quite
different between the PACIAE and AMPT models. There are much more
partons in the AMPT model than in PACIAE. \item The Zhang parton
cascade (ZPC) model \cite{bin} is employed to describe parton
rescattering. In ZPC only elastic scatterings are considered with
$gg\rightarrow gg$ cross section instead of all elastic
interaction cross sections. \item Partons after rescattering are
hadronized by the coalescence model only. \item The dynamics of
the consequent hadronic matter is described by a relativistic
transport (ART) model \cite{li}. In ART the cross sections of
hadron-hardon collisions and hadronic resonances are considered in
detail. Therefore the hadronic rescattering is considered more
carefully in AMPT than in PACIAE.
\end{itemize}
As mentioned in the beginning, the AMPT model has been used
successfully to describe the elliptic flow parameter
\cite{bin1,chen1} in Au+Au collisions at RHIC energy. The AMPT
results given in this paper are calculated by the same code
\cite{chen1} with parameters adjusted to the $v_2$ data. However,
the PACIAE results are calculated with the default parameters.
Thus the two results are not matching each other very well.
Nevertheless, that is irrelevant because we aim to explore the
important physics mentioned above rather than to reproduce the
experimental data. To demonstrate that physics and to estimate its
quantitative importance we used two independent models, both
having a realistic description for multiplicity fluctuations at
the fixed impact parameter, at hand \cite{zhou} to cross check the
correlation between $N_{ch}$ and $b$ and the correlation between
$v_2$ and $b$ ($N_{ch}$).

\vspace{0.5in}
\begin{figure}[htbp]
\epsfig{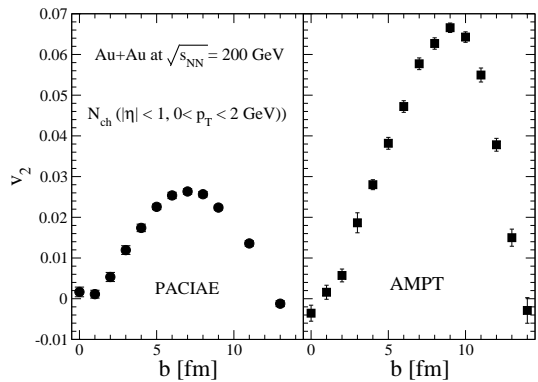} \vspace{-0.1in}
\caption{Impact parameter dependence of integrated $v_2$
calculated by the PACIAE model (left panel) and the AMPT model
(right panel). The error bars indicate the random fluctuations of
$v_2$ at fixed impact parameter.} \label{v2b}
\end{figure}

\section{RESULTS}
The correlation between impact parameter $b$ (in fm) and charged
multiplicity $N_{ch}$ from the PACIAE and AMPT calculations for
Au+Au collisions at $\sqrt{s_{NN}}$=200 GeV are shown in Fig.
\ref{multib}. One sees in the left panel of Fig. \ref{multib} that
$N_{ch}$ is negatively correlated with $b$. A unit of ``fm" change
in impact parameter results in more than 100 charged particle
change in multiplicity. An about 20\% increase in multiplicity
corresponds to about 2 fm decrease in impact parameter. Similar
conclusions can be drawn from the AMPT results in right panel of
Fig. \ref{multib}.

\vspace{0.5in}
\begin{figure}[htbp]
\epsfig{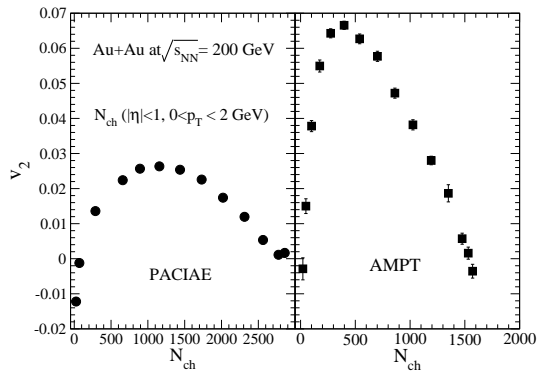} \vspace{-0.1in}
\caption{Charged particle multiplicity dependence of integrated
$v_2$ calculated by the PACIAE model (left panel) and the AMPT
model (right panel). The error bars indicate the random
fluctuations of $v_2$ at fixed charged particle multiplicity.}
\label{v2nch}
\end{figure}

In Fig. \ref{v2b} we give the integrated $v_2$ (event-wise
average) as a function of impact parameter $b$ calculated by the
PACIAE and AMPT models for the Au+Au collisions at
$\sqrt{s_{NN}}$=200 GeV. We see in left panel of Fig. \ref{v2b}
that there is a peak appearing at $b$ nearly equal to the radius
of colliding nucleus. That is a result of competition between the
charge multiplicity and the impact parameter. In central Au+Au
collision the charge multiplicity reaches maximum at zero impact
parameter, where nuclear overlap region is nearly symmetric thus
$v_2$ approaches zero \cite{phob1,phob2}. In the middle peripheral
collision, although the charge multiplicity is going down, the
$v_2$ is large because of the strong asymmetry of nuclear overlap
region (large impact parameter). In the extra peripheral
collision, the asymmetry of nuclear overlap region is very strong,
while the $v_2$ is going down because the multiplicity is too low
to generate pressure gradient. When the impact parameter is around
the radius of colliding nucleus, the charge multiplicity is not so
small, and the asymmetry of nuclear overlap region is considerably
strong, so $v_2$ approaches its maximum. Similar situations can be
seen in the AMPT results shown in the right panel of Fig.
\ref{v2b}. These results are consistent with the PHOBOS and PHENIX
reports about the peak structure in $v_2$ as a function of the
number of participant nucleons ($N_{part}$) in Au+Au collisions at
$\sqrt{s_{NN}}$=200 GeV \cite{phob1,phen2}. Similarly, we give the
integrated $v_2$ as a function of $N_{ch}$ calculated by the
PACIAE and AMPT models for the Au+Au collisions at
$\sqrt{s_{NN}}$=200 GeV in Fig. \ref{v2nch}.

In the left panel of Fig. \ref{multib} we know that a nearly 20\%
increase in multiplicity corresponds to about 2 fm decrease in
impact parameter. This change in impact parameter, in turn,
results in about 0.010 change in $v_2$ (see the sensitive region
of $2<b<7$ in the left panel of Fig. \ref{v2b}). The similar, even
stronger, conclusions can be drawn from the AMPT results in the
right panels of Fig. \ref{multib} and \ref{v2b}. Thus, both the
PACIAE and AMPT models, incorporating random multiplicity
fluctuation, verify that a charge multiplicity bin, which includes
e.g. 20\% of maximum charged multiplicity, spans an impact
parameter range of $\Delta b \sim 2$ fm. That results in a change
of $\Delta v_2 \sim$ 0.010 in the PACIAE calculations. Considering
the maximum $v_2$ in the PACIAE model is just only 0.028, so the
change of $\Delta v_2$ caused by the different impact parameters
in the sample is about 30-40\% of the maximum! This is a very
significant change, which should not be underestimated!

\begin{figure}[ht]
\epsfig{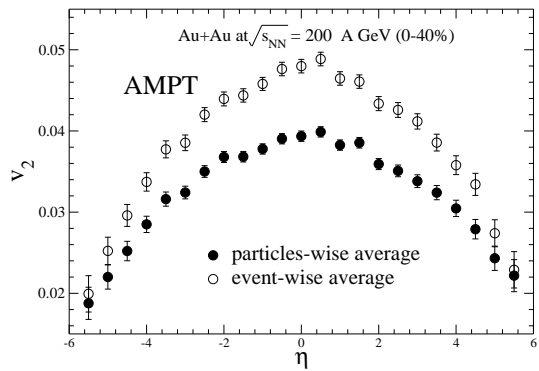}
\vspace{-0.1in} \caption{$v_2$ as a function of $\eta$ calculated
by the AMPT model for $0-40\%$ most central Au+Au at
$\sqrt{s_{NN}}$= 200 GeV, using the two methods of averaging
described in the text.} \label{ampt_eta}
\end{figure}

In \cite{sazh1} the PACIAE model has been used to calculate the
charged hadron $v_2(\eta)$ in 0-40\% most central Au+Au collisions
at $\sqrt {s_{NN}}$=200 GeV by the method of event-wise and
particle-wise averages separately. The $v_2^e(\eta)$ is about 10\%
larger than $v_2^p(\eta)$. This means $\left\langle\overline
{cos(n\phi)}N_{ev}\right\rangle_{ev}$ is smaller than
$\left\langle \overline{cos(n\phi)}\right\rangle_{ev}\left\langle
N_{ev} \right \rangle_{ev}$ and demonstrates the negative
correlation between $\overline{cos(n\phi)}$ and $N_{ev}$. Here we
use the AMPT model \cite{ampt1} to calculate $v_2^e(\eta)$ and
$v_2^p(\eta)$ in 0-40\% most central Au+Au collisions at $\sqrt
{s_{NN}}$=200 GeV again. The AMPT results of $v_2^e(\eta)$ are
nearly 20\% larger than $v_2^p(\eta)$, as shown in Fig.
\ref{ampt_eta}.

\vspace{0.5in}

\section{CONCLUSIONS}
In summary, we have used the parton and hadron cascade model
PACIAE and a multiple phase transport model AMPT with string
melting to investigate the elliptic flow parameter $v_2$ by taking
into account the event-by-event charge multiplicity  fluctuations
and to cross check the physics of initial spatial asymmetry, final
momentum asymmetry, and their fluctuations and correlations. The
correlations between the charge multiplicity and impact parameter
as well as between the elliptic flow parameter and impact
parameter (charge multiplicity) are calculated.

It turned out that the charge multiplicity is negatively
correlated with impact parameter. The $v_2$ as a function of
impact parameter first increases with increasing $b$, reaches its
maximum at $b$ close to the radius of colliding nucleus, and turns
to decrease with increasing $b$ at last. This is because the $v_2$
is determined by two driving forces, multiplicity and impact
parameter.

The averaging procedure in the definition of elliptic flow
parameter is reexamined by the AMPT calculations for $v_2^e(\eta)$
and $v_2^p(\eta)$ in 0-40\% most central Au+Au collisions at
$\sqrt{s_{NN}}$=200 GeV. The AMPT results of $v_2^e(\eta)$ are
about 20\% larger than $v_2^p(\eta)$, which are larger than the
corresponding PACIAE results \cite{sazh1}. This emphasizes again
the necessity of theoretically calculating the elliptic flow
parameter $v_2$ by the event-wise average, in order to be
consistent with the experimental extraction of $v_2$, which is
really event-wise average \cite{sazh1,casa}.

\acknowledgments
Financial support from NSFC, China (grants 10705012, 10475032,
10605040, 10635020, 10975062,) RCN, Norway (grants 101937 and 171247),
the Chinese-Norwegian RCN-NFSC collaboration grant, no. 10710886, and
Programme of Introducing Talents of Discipline to Universitiesnder
under Grant No. B08033 are gratefully acknowledged.

\end{document}